\documentclass[12pt]{emulateapj}
\usepackage{amsfonts}
\usepackage{bigstrut,array}

\begin{document}

\def\simeq{
\mathrel{\raise.3ex\hbox{$\sim$}\mkern-14mu\lower0.4ex\hbox{$-$}}
}


\def\lsun{{\rm L_{\odot}}}
\def\msun{{\rm M_{\odot}}}
\def\rsun{{\rm R_{\odot}}} 
\def\lta{\la}
\def\gta{\ga}
\def\be{\begin{equation}}
\def\ee{\end{equation}}
\def\lsun{{\rm L_{\odot}}}
\def\le{{L_{\rm Edd}}}
\def\msun{{\rm M_{\odot}}}
\def\rsun{{\rm R_{\odot}}}
\def\rp{{R_{\rm ph}}}
\def\rs{{R_{\rm s}}}
\def\mw{{\dot M_{\rm w}}}
\def\mo{{\dot M_{\rm out}}}
\def\me{{\dot M_{\rm Edd}}}
\def\tc{{t_{\rm C}}}
\def\rc{{R_{\rm core}}}
\def\mc{{M_{\rm core}}}
\def\mbh{{M_{\rm BH}}}
\def\e{{\dot m_{\rm E}}}

\title{Clearing Out a Galaxy}

\author{ Kastytis~Zubovas\altaffilmark{1} and Andrew~King\altaffilmark{1}}

\altaffiltext{1} {Theoretical Astrophysics Group, University of
Leicester, Leicester LE1 7RH, U.K.; ark@astro.le.ac.uk}

\begin{abstract}

It is widely suspected that AGN activity ultimately sweeps galaxies clear of
their gas. We work out the observable properties required to achieve
this. Large--scale AGN--driven outflows should have kinetic luminosities $\sim
\eta\le/2 \sim 0.05\le$ and momentum rates $\sim 20\le/c$, where $\le$ is the
Eddington luminosity of the central black hole and $\eta\sim 0.1$ its
radiative accretion efficiency. This creates an expanding two--phase medium in
which molecular species coexist with hot gas, which can persist after the
central AGN has switched off. This picture predicts outflow velocities $\sim
1000 - 1500$~km\,s$^{-1}$ and mass outflow rates up to $4000~\msun\,{\rm
  yr}^{-1}$ on kpc scales, fixed mainly by the host galaxy velocity dispersion
(or equivalently black hole mass). All these features agree with those of
outflows observed in galaxies such as Mrk231. This strongly suggests that AGN
activity is what sweeps galaxies clear of their gas on a dynamical timescale
and makes them red and dead. We suggest future observational tests of this
picture.

\end{abstract}


\section{Introduction}

Recently, three groups \citep{Feruglio2010A&A, Rupke2011ApJ, Sturm2011ApJ}
have used molecular spectral line observations to reveal fast ($v_{\rm out}
\sim 1000$~km\, s$^{-1}$) kpc--scale, massive ($\mo \sim 1000~\msun\, {\rm
  yr}^{-1}$) outflows in the nearby quasar Mrk231. Other galaxies show
indications of similar phenomena \citep[e.g.][]{Riffel2011MNRAS,
  Riffel2011MNRASb, Sturm2011ApJ}. These appear to show how quasar feedback
can transform young, star--forming galaxies into red and dead spheroids. All
three groups reach this conclusion for Mrk231 essentially by noting that the
mass outflow rate $\dot M_{\rm out}$ and the kinetic energy rate $\dot E_{\rm
  out} = \dot M_{\rm out}v_{\rm out}^2/2$ of the outflow are too large to be
driven by star formation, but comparable with those predicted in numerical
simulations of AGN feedback. The kinetic energy rate is a few percent of the
likely Eddington luminosity $\le = 4\pi GM c/\kappa$ of the central black
hole, of mass $M$ (where $\kappa$ is the electron--scattering opacity). The
outflowing material must have a multi-phase structure, because $v_{\rm out}$
greatly exceeds the velocity corresponding to the molecular dissociation
temperature ($v_{\rm diss} \lesssim 10$ km~s$^{-1}$; see Section 5 below).

In a recent paper \citep{King2011MNRAS} we showed that large--scale flows of
this type (technically, an energy--driven flow, see Section 3) can indeed
drive much of the interstellar gas out of a galaxy bulge on a dynamical
timescale $\sim 10^8$~yr, leaving it red and dead, provided that the central
supermassive black hole accretes for about twice the Salpeter time after
reaching the value set by the $M - \sigma$ relation.  In
\citet{Power2011MNRAS} we showed that the remaining bulge mass is close to the
value set by the observed black--hole -- bulge--mass relation
\citep[e.g.][]{Haering2004ApJ}. However we did not investigate the observable
features of this process, including in particular the way that the
interstellar gas is swept up.

We return to this problem here, as it offers a clear observational test of the
idea that AGN outflows are responsible for making galaxies red and dead. To
keep our treatment as general as possible (specifically, independent of the
details of numerical simulations) we adopt a simple analytic approach. We find
that this process predicts outflow velocities $\sim 1000-1500$~km\,s$^{-1}$,
and mass outflow rates up to $\sim 4000~\msun\,{\rm yr}^{-1}$, several
hundred times the Eddington value, in good agreement with observations. In
addition, we find that the observable momentum outflow rate is $\sim 20$ times
greater than $L/c$ of the driving AGN, also in agreement with observations. We
conclude that AGN outflows are good candidates for the agency sweeping
galaxies clear of gas.

\begin{figure*}
  \centering
    \includegraphics[width=\textwidth]{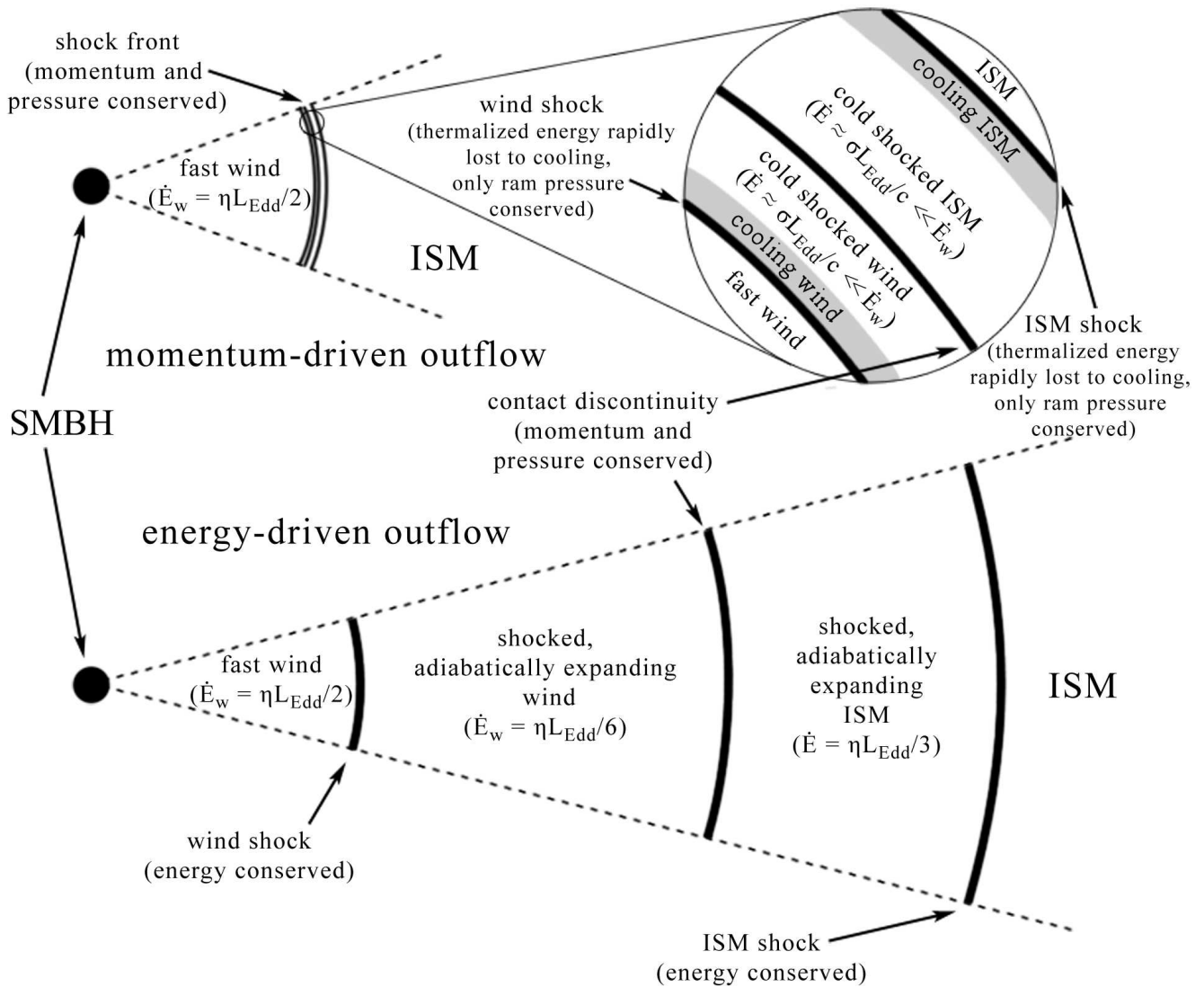}
  \caption{Diagram of momentum--driven (top) and energy--driven
    (bottom) outflows. In both cases a fast wind (velocity $\sim \eta
    c \sim 0.1c$) impacts the interstellar gas of the host galaxy,
    producing an inner reverse shock slowing the wind, and an outer
    forward shock accelerating the swept--up gas. In the
    momentum--driven case, the shocks are very narrow and rapidly cool
    to become effectively isothermal. Only the ram pressure is
    communicated to the outflow, leading to very low kinetic energy
    $\sim (\sigma/c)\le$. In an energy--driven outflow, the shocked
    regions are much wider and do not cool. They expand adiabatically,
    communicating most of the kinetic energy of the wind to the
    outflow (in simple cases approximately 1/3rd is retained by the
    shocked wind). The outflow radial momentum flux is therefore
    greater than that of the wind. Momentum--driven flows occur when
    shocks happen within $\sim 1$~kpc of the AGN, and establish the $M
    - \sigma$ relation \citep{King2003ApJ, King2005ApJ}. Once the
    supermassive black hole mass attains the critical $M - \sigma$
    value, the shocks move further from the AGN and the outflow
    becomes energy--driven. This produces the observed large--scale
    flows, which probably sweep the galaxy clear of gas.}
  \label{fig:outflow}
\end{figure*}

\section{Winds}

To drive outflows with $\dot E_{\rm out}$ approaching $\le$ at large radius,
the active nucleus of a galaxy must somehow communicate this luminosity from
its immediate vicinity. Direct transport by radiation is problematic, not
least because galaxies are generally optically thin \citep[However, dust
  opacity may be large enough to absorb this radiation and provide feedback;
  see][]{Murray2005ApJ}. Jets are sometimes invoked, but are relatively
inefficient because they tend to drill holes in the interstellar medium rather
than driving it bodily away. Accordingly the most likely connection is via
high--velocity wide--angle winds expelled from the vicinity of the nucleus by
radiation pressure \citep[e.g.][]{Pounds2003MNRASb, Pounds2003MNRASa}. Recent
observations suggest that such winds are very common in AGN
\citep{Tombesi2010A&A, Tombesi2010ApJ}. In this paper, we use the term `wind'
to refer to the mildly relativistic ($v \sim 0.1c$) ejection of accretion disc
gas from the immediate vicinity of the SMBH resulting from Eddington
accretion, and `outflow' (see Section 4) for the large--scale nonrelativistic
flows caused by the interaction between the wind and the galaxy's ambient gas.

The winds have simple properties. With mass rate $\mw \sim \me$, where $\me =
\le/\eta c^2$ is the Eddington accretion rate and $\eta$ is the accretion
efficiency, a wide--angle wind has scattering optical depth $\sim 1$
\citep[][]{King2003MNRASb}, assuming that the covering factor of absorbing gas
is close to unity (see the discussion below eqn \ref{v}).  So each driving
photon on average scatters about once before escaping to infinity and gives up
all of its momentum to the wind, so that the wind mass flow rate $\mw$ and
velocity $v$ obey
\begin{equation}
\mw v \sim {L\over c}
\label{mom}
\end{equation}
Defining $\dot m = \mw/\me \sim 1$ as the Eddington factor of the wind, we
immediately find
\begin{equation}\label{v}
{v\over c} \sim {\eta\over \dot m} \sim 0.1
\end{equation}
\citep[cf][]{King2010MNRASa}. The likely ionization equilibrium of the wind is
such that it produces X--rays \citep{King2010MNRASa}. In line with these
expectations, blueshifted X--ray iron absorption lines corresponding to
velocities $\sim 0.1c$ are seen in a significant fraction of local AGN
\citep[e.g.][]{Pounds2003MNRASb, Pounds2003MNRASa, Tombesi2010A&A,
  Tombesi2010ApJ}, justifying our assumption of a covering factor close to
unity. In all cases the inferred wind mass flow rates agree with
eqn. (\ref{mom}). So these black hole winds have momentum and energy rates
\begin{equation}
\dot P_w \sim \dot m{\le\over c} \sim {\le\over c},
\label{pdot}
\end{equation}
and
\begin{equation}
\dot E_w = {1\over 2}\mw v^2 \sim {\eta\over 2\dot m}\le \sim 0.05\le,
\label{edot}
\end{equation}
where we have used eqns (\ref{mom}, \ref{v}) in eqn (\ref{edot}).

\section{Shocks}

The expression (\ref{edot}) for the energy rate of a black hole wind is
obviously promising for driving the observed large--scale outflows. Although
the interstellar medium is clumpy, the outflow bubble inflated by the wind
easily sweeps past the clumps, affecting the diffuse gas
\citep[e.g.][]{MacLow1988ApJ}. Furthermore, the clouds are shocked by the
passing outflow and evaporate inside the hot wind bubble \citep{Cowie1977ApJ},
so most of their material also joins the outflow. A detailed treatment of the
interaction between the clumpy ISM and the wind--driven outflow is beyond the
scope of this paper, although we address some of the implications in Section
6. In the present analysis we assume that most of the sightlines from the SMBH
are covered with diffuse medium, irrespective of whether they are also
obscured by clumps.

The question now is how efficiently the wind energy is transmitted to
the outflow. This depends crucially on how the wind interacts with the
diffuse interstellar medium of the host galaxy. Since the wind is
hypersonic, it must decelerate violently in a reverse shock, and
simultaneously drive a forward shock into the host ISM. There are two
possible outcomes, which are realised under different conditions in
galaxies.

The first outcome (momentum--driven flow) occurs if the shocked wind gas can
cool on a timescale short compared with the motion of the shock pattern. In
this case the shocked wind gas is compressed to high density and radiates away
almost all of the wind kinetic energy (i.e. $\dot{E}_{\rm out} << \dot{E}_w =
(\eta/2\dot m)\le$). This shocked wind has gas pressure equal to the
pre--shock ram pressure $\dot P_{\rm w} \simeq \le\dot m /c \propto M$, and
this pushes into the host ISM.

The second case (energy--driven flow) occurs if the shocked wind
gas is not efficiently cooled, and instead expands as a hot bubble. Then the
flow is essentially adiabatic, and has the wind energy rate, i.e. $\dot E_{\rm
  out} \simeq \dot E_{\rm w} = (\eta/2\dot m)\le \sim 0.05\le$ (from eqn
\ref{edot}). The hot bubble's thermal expansion makes the driving into the
host ISM more vigorous than in the momentum--driven case.  Observed
galaxy--wide molecular outflows must be energy--driven, as demonstrated
directly by their kinetic energy content (cf eqn \ref{edot}).

Which of these two very different cases occurs at a given point depends on the
cooling of the shocked gas. It is easy to show that the usual atomic cooling
processes (free--free and free--bound radiation) are negligible in all
cases. The dominant process tending to cool the shocked black hole wind is the
inverse Compton effect \citep{Ciotti1997ApJ}. The quasar radiation field is
much cooler than the wind shock temperature (typically $\sim 10^7$~K and $\sim
10^{11}$~K respectively), and so cools the shocked wind provided that it is
not too diluted by distance. Equations 8 and 9 of \citet{King2003ApJ} show
that this holds if and only if the shock is at distances $R \la 1$~kpc from
the active nucleus, since the Compton cooling time goes as $R^2$ and the flow
time typically as $R$. So we expect momentum--driven flow close to the
nucleus, and energy--driven flow if gas can be driven far away from it.

We note that \citet{Silk2010ApJ} claim that an energy--driven flow never
occurs. However they seem to have considered the cooling of the ambient gas,
rather than the shocked wind and swept--up ISM: the cooling function they
use extends only to temperatures $\sim 10^7$~K, rather than the wind shock
temperature $\sim 10^{11}$~K -- see \citet{King2011MNRAS} for details.

\section{Outflows}

For an isothermal ISM density distribution with velocity dispersion $\sigma$
and gas fraction $f_c$ (the ratio of gas density to background potential
density) one can solve analytically the equation of motion for the shock
pattern for both momentum--driven flow \citep{King2003ApJ, King2005ApJ} and
energy--driven flow \citep{King2005ApJ, King2011MNRAS}.

In the momentum--driven case there are two distinct flow patterns, depending
on the black hole mass $M$. For $M < M_{\sigma}$, where
\begin{equation}
\label{msig}
M_{\sigma} = {f_c\kappa\over \pi G^2}\sigma^4 \simeq 4\times
10^8\msun\sigma_{200}^4
\end{equation}
(with $f_c = 0.16$ (its cosmological value) and $\sigma_{200} =
\sigma/(200~{\rm km\,s^{-1}})$) the wind momentum is too weak to drive away
the swept--up ISM, and the flow stalls at some point. For $M > M_{\sigma}$ the
wind momentum drives the swept--up matter far from the nucleus. It is
intuitively reasonable to assume that the black hole cannot easily grow its
mass significantly beyond the point where it expels the local interstellar gas
in this way, i.e. beyond $M_{\sigma}$. Equation (\ref{msig}) is very close to
the observed $M - \sigma$ relation, despite having no free parameter.
Detailed calculations \citep{Power2011MNRAS} show that the SMBH is likely to
grow for 1-2 additional Salpeter times after it reaches $M_{\sigma}$,
increasing its final mass by a factor of a few. This process is even more
pronounced at higher redshift, as then it takes longer for the outflow to
clear the galaxy, so the SMBH must be active for longer.

We conclude that outflows drive gas far from the nucleus, and thus become
energy--driven, once $M \gtrsim M_{\sigma}$.  This is evidently the case
needed to explain the molecular outflows seen in Mrk231 and other galaxies.

\begin{table*}
\centering

  \caption{Observed outflow parameters}

  \setlength{\extrarowheight}{1.5pt}

  \begin{tabular}{c | c c c c c }
    \hline \hline 
    Object & $M_{\rm BH} / \msun$ & $\sigma /$ km s$^{-1}$ &
    $L_{\rm bol} /$~erg s$^{-1}$ ($l$) & $\mo / \msun$ yr$^{-1}$ & $v_{\rm
      out} /$km s$^{-1}$ \bigstrut \\

    \hline 

    Mrk231$^{(a)}$ & $4.7 \cdot 10^{7(b)}$ & $120^{(b)}$ & $45.69^{(c)} ~
    (0.80)$ & $420$ & $1100$  \bigstrut[t] \\

    Mrk231$^{(d)}$ & $4.7 \cdot 10^{7}$ & $120$ & $45.69 ~ (0.80)$ & $700$ &
    $750$ \\

    Mrk231$^{(e)}$ & $4.7 \cdot 10^{7}$ & $120$ & $46.04^{(f)} ~ (1.8)$ &
    $1200$ & $1200$ \\

    IRAS 08572+3915$^{(e)}$ & $\sim 4.5 \cdot 10^{7*}$ & $120^{***}$ &
    $45.66 ~ (1^*)$ & $970$ & $1260$ \\

    IRAS 13120--5453$^{(e)}$ & $5.3 \cdot 10^{6*}$ & $70^{***}$ & $44.83~(1^*)$ &
    $130$ & $860$ \\

    IRAS 17208--0014$^{(e)**}$ & $-$ & $-$ & $45.11~(\ll 1)$ & $90$ & $370$ \\

    Mrk1157$^{(g)}$ & $8.3 \cdot 10^6$ & $100$ & $42.57~(3.4 \cdot 10^{-3})$ &
    $6$ & $350$ \\

    2QZJ002830.4-281706$^{(h)}$ & $5.1 \cdot 10^{9(i)}$ & $385^{***}$
    & $46.58~(5.8 \cdot 10^{-2})$ & $2000$ & $2000$ \\

    \hline
    \hline
  \end{tabular}

\begin{list}{}{}
\item[ ] {\footnotesize Outflows observed in molecular gas (Mrk231,
  IRAS 08572+3915, IRAS 13120--5453 and IRAS 17208--0014), and warm
  ionised gas (Mrk1157, 2QZJ002830.4-281706). $^{*}$ - the AGN is
  assumed to be radiating at its Eddington limit; $^{**}$ - the galaxy
  is starburst--dominated \citep{Riffel2011MNRAS}, so we expect a low
  Eddington factor and hence make no estimates; $^{***}$ - the SMBH is
  assumed to lie on the $M-\sigma$ relation; note that this may be
  questionable in some cases \citep[cf][]{McConnell2011Natur}.}
\end{list}

\begin{list}{}{}
\item[ ] {\footnotesize    References: $^a$ - \citet{Rupke2011ApJ};
  $^b$ - \citet{Tacconi2002ApJ};
  $^c$ - \citet{Lonsdale2003ApJ};
  $^d$ - \citet{Feruglio2010A&A};
  $^e$ - \citet{Sturm2011ApJ};
  $^f$ - \citet{Veilleux2009ApJS};
  $^g$ - \citet{Riffel2011MNRAS};
  $^h$ - \citet{CanoDiaz2011arXiv};
  $^i$ - \citet{Shemmer2004ApJ}.}
\end{list}

  \label{table:obs}
\end{table*}

\begin{table*}
\centering

  \setlength{\extrarowheight}{1.5pt}

  \caption{Observationally derived versus theoretically predicted
    outflow parameters}

  \begin{tabular}{c | c c c c c c}

    \hline \hline 
    Object & $\frac{\dot{E}_{\rm out}}{0.05 L_{\rm bol}}$ &
    $\frac{\mo v_{\rm out} c}{L_{\rm bol}}$ & $f_{\rm L} \equiv
    \frac{\mo}{\dot{M}_{\rm acc}}$ & $f_{\rm L, pred.}$ & 
    $\dot{M}_{\rm pred.} / \msun$ yr$^{-1}$ 
    & $v_{\rm pred.} /$~km s$^{-1}$ \bigstrut \\ 

    \hline 

    Mrk231 & $0.66$ & $18$ & $490 = 22^2$ & $840$ & $880$ & $810$ \bigstrut[t] \\

    Mrk231 & $0.51$ & $20$ & $820 = 29^2$ & $840$ & $880$ & $810$ \\

    Mrk231 & $1.0$ & $25$ & $1400 = 37^2$ & $1110$ & $1150$ & $1060$ \\

    IRAS 08572+3915 & $2.1$ & $50$ & $1200 = 35^2$ & $910$ & $950$ & $875$ \\

    IRAS 13120--5453 & $0.88$ & $31$ & $1080 = 33^2$ & $1870$ & $220$ & $610$ \\

    IRAS 17208--0014 & $0.06$ & $4.9$ & $396 = 20^2$ & $-$ & $-$ & $-$ \\

    Mrk1157 & $1.3$ & $110$ & $9270 = 96^2$ & $170$ & $85$ & $115$ \\

    2QZJ002830.4-281706 & $1.3$ & $20$ & $307 = 17.5^2$ & $74$ & $8200$ & $740$ \\

    \hline
    \hline
  \end{tabular}

\begin{list}{}{}
\item[ ] {\footnotesize The first three columns give quantities
  derived from observations of large--scale outflows summarized in
  Table \ref{table:obs}.  The last three columns give the
  mass--loading parameter, mass sweep--out rate and terminal velocity
  predicted by our equations (\ref{load2}), (\ref{out}) and
  (\ref{vout}) respectively. We assume the simplest case of an
  isotropic outflow. Collimation would reduce the predicted mass
  outflow rate and increase the predicted outflow velocity. With one
  outlier (see below), the outflow kinetic energy is always very close
  to $5\%$ of $L_{\rm bol}$ (1st column) as predicted by
  eq. (\ref{edot}), and the momentum loading (2nd column) is always
  very similar to the square root of the mass loading (rhs of 3rd
  column), as predicted by eq. (\ref{eq:dotp}). It is striking that
  the relation holds for local quasars (Mrk231), high-redshift quasars
  (2QZJ002830.4-281706) and low luminosity galaxies (Mrk1157).  The
  last two columns can be directly compared with the last two columns
  of Table \ref{table:obs}; the discrepancies arise due to strong
  outflow collimation. The only significant outlier, IRAS 17208--0014,
  is known to be a starburst-dominated galaxy, so we would not expect
  the outflow to be dominated by the AGN contribution.}
\end{list}

  \label{table:pred}
\end{table*}

\section{Large--Scale Flows}

In an energy--driven flow the adiabatic expansion of the shocked wind pushes
the swept--up interstellar medium in a `snowplow'. \citet{King2011MNRAS}
derive the analytic solution for the expansion of the shocked wind in a galaxy
bulge with an isothermal mass distribution. With AGN luminosity $l\le$, all
such solutions tend to an attractor
\begin{equation}
\dot R = v_e \simeq \biggl[{2\eta lf_c\over 3f_g}\sigma^2c\biggr]^{1/3} \simeq
925\sigma_{200}^{2/3}(lf_c/f_g)^{1/3}~{\rm km\ s}^{-1}
\label{ve}
\end{equation} 
until the central AGN luminosity decreases significantly at some radius $R =
R_0$, when the expansion speed decays as
\begin{equation}
\dot R^2 = 3\biggl(v_e^2 + {10\over 3}\sigma^2\biggr)\biggl({1\over
  x^2} - {2\over 3x^3}\biggr) - {10\over 3}\sigma^2
\label{dotr}
\end{equation}
where $x = R/R_0\geq 1$. In eq. (\ref{ve}), $f_g$ is the gas fraction relative
to all matter. This may be lower than the value $f_c$ prevailing when the
earlier momentum--driven outflow establishes the $M - \sigma$ relation
(\ref{msig}), as gas may be depleted through star formation for example.

The solutions (\ref{ve}, \ref{dotr}) describe the motion of the contact
discontinuity where the shocked wind encounters swept--up interstellar gas
(see Figure \ref{fig:outflow}). The observed molecular lines are likely to
come from the shocked interstellar gas ahead of this discontinuity -- its
temperature is much lower ($\sim 10^7$~K) than that of the shocked wind, as we
shall see. The outer shock must run ahead of the contact discontinuity into
the ambient interstellar medium in such a way that the velocity jump across it
is a factor $(\gamma +1)/(\gamma -1)$ (where $\gamma$ is the specific heat
ratio). This fixes its velocity as
\begin{equation}
v_{\rm out} = {\gamma + 1\over 2}\dot R \simeq
1230\sigma_{200}^{2/3}\left({lf_c\over f_g}\right)^{1/3}~{\rm km\ s}^{-1}
\label{vout}
\end{equation}
(where we have used $\gamma= 5/3$ in the last form). This corresponds to a
shock temperature of order $10^7$~K for the forward shock into the
interstellar medium (as opposed to $\sim 10^{10 - 11}~K$ for the wind
shock). Since the outer shock and the contact discontinuity are very close
together when energy--driven flow starts (see Fig. \ref{fig:outflow}) this
means that the outer shock is always at
\begin{equation}
R_{\rm out} = {\gamma + 1\over 2}R.
\end{equation}
The outflow rate of shocked interstellar gas is
\begin{equation} 
\dot{M}_{\rm out} = {{\rm d}M(R_{\rm out})\over {\rm d}t} = {(\gamma +
  1)f_{\rm g} \sigma^2\over G}\dot R.
\end{equation} 
Assuming $M = M_{\sigma}$, the wind outflow rate is
\begin{equation}
\dot{M}_{\rm w} \equiv \dot{m}\dot{M}_{\rm Edd} = \frac{4 f_{\rm c} \dot{m}
  \sigma^4}{\eta c G}.
\end{equation}
We can now define a mass--loading factor for the outflow, which is the
  ratio of the mass flow rate in the shocked ISM to that in the wind:
\begin{equation}
f_{\rm L} \equiv {\mo\over\mw} = {\eta(\gamma + 1)\over 4\dot m}{f_g\over
  f_c}{\dot Rc\over \sigma^2}.
\label{load}
\end{equation}
Then the mass outflow rate is
\begin{equation}
\mo = f_{\rm L}\mw = {\eta(\gamma + 1)\over 4}{f_g\over f_c}{\dot Rc\over
  \sigma^2}\me.
\end{equation}
If the AGN is still radiating at a luminosity close to Eddington, we
have $\dot R = v_e$, and using (\ref{ve}) gives
\begin{equation}
f_{\rm L} = \left({2\eta c\over 3\sigma}\right)^{4/3}\left({f_g\over
  f_c}\right)^{2/3}{l^{1/3}\over \dot m} \simeq
460\sigma_{200}^{-4/3}{l^{1/3}\over \dot m},
\label{load2}
\end{equation}
and
\begin{equation}
\mo \simeq 3700\sigma_{200}^{8/3}l^{1/3}~\msun\,{\rm yr}^{-1}
\label{out}
\end{equation}
for typical parameters, $f_{\rm g} = f_{\rm c}$ and $\gamma = 5/3$. If the
central quasar is no longer active, the mass outflow rate evidently declines
as $\dot R/v_e$ times this expression, with $\dot R$ given by (\ref{dotr}).

It is easy to check from (\ref{vout}, \ref{out}) that the approximate equality
\begin{equation} 
\frac{1}{2}\dot{M}_{\rm w} v_{\rm w}^2 \simeq
\frac{1}{2}\dot{M}_{\rm out} v_{\rm out}^2.  
\label{energy}
\end{equation} 
holds, i.e. most of the wind kinetic energy ultimately goes into the
mechanical energy of the outflow, as expected for energy driving. One can show
from the equations in \citet{King2005ApJ} that if the quasar is still active
and accreting close to its Eddington rate, the shocked wind retains
$1/3$ of the total incident wind kinetic energy $\dot M_{\rm w}v_{\rm w}^2/2$,
giving $2/3$ to the swept--up gas. The energy retained in the wind and the
swept--up gas have potentially observable emission signatures (see
Discussion).

Equation (\ref{energy}) means that the swept--up gas must have a momentum rate
greater than the Eddington value $\le/c$, since we can rewrite it as
\begin{equation} 
\frac{\dot{P}_{\rm w}^2}{2 \dot{M}_{\rm w}} \simeq \frac{\dot{P}_{\rm
    out}^2}{2 \dot{M}_{\rm out}}, 
\end{equation}
where the $\dot P$'s are the momentum fluxes. With $\dot{P}_{\rm w} = L_{\rm
  Edd}/c$, we have
\begin{equation} \label{eq:dotp}
\dot{P}_{\rm out} = \dot{P}_{\rm w} \left(\frac{\dot{M}_{\rm
    out}}{\dot{M}_{\rm w}}\right)^{1/2} = \frac{L_{\rm Edd}}{c} f_{\rm
  L}^{1/2} \sim 20 \sigma_{200}^{-2/3} l^{1/6} {\le\over c}
\end{equation} 
where $f_{\rm L}$ is the mass loading factor of the outflow. The factor
$f_{\rm L}^{1/2} \sim 20$ is the reason why observations consistently show
$\dot{M}_{\rm out} v_{\rm out} > L_{\rm Edd}/c$.

\section{Discussion}

We have shown that large--scale outflows driven by wide--angle AGN winds
should have typical velocities $v_{\rm out} \sim 1000 - 1500$~km\,s$^{-1}$ and
mass flow rates up to $\dot M_{\rm out} \sim 4000~\msun\,{\rm yr}^{-1}$ (eqns
\ref{vout}, \ref{out}) if the central quasar is still active, with lower
values if it has become fainter. Our equations (\ref{vout}, \ref{out})
directly relate the outflow velocities and mass rates to the properties of the
host galaxy. The outflows should have mechanical luminosities $\dot E_{\rm
  out} \sim (\eta/2)\le \sim 0.05\le$, but (scalar) momentum rates $\dot P
_{\rm out}\sim 20\le/c$.  These predictions agree well with observations (see
Tables \ref{table:obs} and \ref{table:pred}). We conclude that AGN outflows
may well be what sweeps galaxies clear of gas.

Our picture predicts several other features that may aid in
interpreting observations. It suggests that the molecular outflows come from
clumps of cool gas embedded in the outflowing shocked ISM. They are entrained
by the advancing outer shock front and persist for a long time. We note that
this shock front is Rayleigh--Taylor stable since interstellar gas is
compressed here. Further, the temperature of the shocked ISM is in the right
range for thermal instability \citep{McKee1977ApJ}, and Richtmyer-Meshkov
instabilities \citep{Kane1999ApJ} induced by the forward shock mean that new
cold clumps may form in the outflow behind it. Simulations by Zubovas \&
Nayakshin (in preparation) show that up to $10\%$ of the total mass in the
outflow may be locked up in this cold phase. This agrees with the conversion
factor of $\sim10\%$ used in the papers cited in Table \ref{table:obs} to
estimate the total mass outflow rate from observed molecular species. Our
model therefore predicts both the total mass outflow rates (eqn \ref{out}) and
the observational signatures used to estimate them, in good agreement with
observation (Table \ref{table:pred}).

The inner wind shock presumably accelerates cosmic ray particles, and gamma
rays result when these hit the colder ISM and shocked wind. The outflows are
therefore directly comparable with the gamma--ray emitting bubbles in our
Galaxy recently discovered by {\it Fermi} \citep{Su2010ApJ}. One can interpret
these as relics of the Milky Way's last quasar outburst about 6~Myr ago by
noting that the greater density of the Galactic plane must pinch a
quasi--spherical quasar outflow into a bipolar shape \citep{Su2010ApJ,
  Zubovas2011MNRAS}. The gamma--ray emission from distant galaxies discussed
here should be intrinsically stronger than in the Milky Way, but the long
integration time required to detect the Galactic bubbles means that these
outflows may be undetectable with current instruments.

Perhaps more promisingly, these cosmic ray electrons cool and emit
synchrotron radiation in the radio band. This radiation may be
observable and so it would be interesting to check whether there are
kpc or sub--kpc scale radio bubbles associated with the outflows.

\acknowledgments
 
We thank Sylvain Veilleux and David Rupke for helpful discussions, and
the referee for a very thoughtful and helpful report. Research in
theoretical astrophysics at Leicester is supported by an STFC Rolling
Grant. KZ is supported by an STFC research studentship.


\end{document}